# A Novel Methodologyof Router-To-ASMapping inspired by Community Discovery

Liu Weiyi, Jiang Qing, Fei Gaolei, Yuan Mingkai and Hu Guangmin
School of Communication and Information Engineering
University of Electronic Science and Technology of China (UESTC)
Chengdu, Sichuan, China
unique_liu@163.com, fgl@uestc.edu,cn, mingkaiyuan@126.com, hgm@uestc.edu.cn

***Abstract*—In the last decade many works has been done on the Internet topology at router or autonomous system (AS) level. As routers is the essential composition of ASes while ASes dominate the behavior of their routers. It is no doubt that identifying the affiliation between routers and ASes can let us gain a deeper understanding on the topology. However, the existing methods that assign a router to an AS just based on the origin ASes of its IP addresses, which does not make full use of information in our hand. In this paper, we propose a methodology to assign routers to their owner ASes based on community discovery tech. First, we use the origin ASes information along with router-pairs similarities to construct a weighted router level topology; secondly, for enormous topology data (more than 2M nodes and 19M edges) from CAIDA ITDK project, we propose a fast hierarchy clustering which time and space complex are both linear to do ASes community discovery, last we do router-to-AS mapping based on these ASes communities. Experiments show that combining with ASes communities our methodology discovers, the best accuracy rate of router-to-AS mapping can reach to 82.62%, which is drastically high comparing to prior works that stagnate on 65.44%.**

## I. Introduction

As a hot topic in the networking research group, the topology of the Internet drawn a considerable attention of many researchers. They probe and construct the topology at router or AS level [1]-[4], then use it to study the characteristics of the topology, design topology generators, develop network protocols and evaluate their performance. The classic definition of an Autonomous System (AS) is a set of routers under a single technical administration [5], which means routers are the foundation of an AS and all of the routers in an AS observe the same routing policy decided by the AS. Thus ASes can affect the behavior of their routers while routers can reflect some features of their owner ASes. In addition, router level topology can provide a more detail intra and inter connection of ASes. Thus it is necessary to merge the router and the AS level topology, i.e., mapping each router to the AS that own it, which can help us deepen the understanding of the topology. For example, through the mapping we can obtain an accurate AS-level traceroute tool [6], and study the correlation between the degree of the ASes and the number of their routers [7].

However, there are only a few works focus on this issue. To complete their studies, Tangmunarunkit et al.[7][8]was the first to try to solve this problem. They directly map the router to the AS that appears most frequently on the origin ASes of the router. In addition, Pansiot et al. [9] proposed five simple rules that include two probabilistic rules and three empirical rules to identify the owner ASes of the AS border routers (ASBR), such like global election which is same with the method used by Tangmunarunkit et al. Both of them have not validate the accuracy of their method. As far as we know, still now, the only method that have been validated was proposed in [10]. Huffaker et al. design five router assignment heuristics based on the origin ASes of routers, and validated them on the ground truth provided by two ISPs and five research networks. They also tested all combinations of pairs of heuristics, and found out the most successful pair that is Election + Degree. Throughout these works, they only take into account the information from router itself or its neighbors, which is clearly limited.

Because various types of real-world complex networks (include router level topology) have certain community structure [11], where a community refers to a set of vertices that behave differently from the rest of the system [12], and may has a dense connection within it [13][14]. According to the definition of AS [15], AS is in the same collection of technical management of a group of routers. Such definition makes router within the same AS may have more connection, while others have less. It seems that ASes behavior from Internet is identical to communities from complex network. In order to enhance the community attribute of router level topology. Correlativity is be introduced to depict router pair relationships, because although routers may not have dense connections under an AS, their correlational relationships cannot vary differently. Inspired by this, we use original AS information of IP interfaces to calculate node similarity which can quantify router pair correlations. Based on the router pair correlations, we construct a weighted graph where routers as vertices and weights stand for correlations. Then we implement community discovery methods on the weighted router level topology.

We think community discovery methods can be used upon this graph to find underlying ASes communities, and router-to-AS mapping can leverage these communities' information to be achieved instead of through old fashion ways described in the first place. Experiments results in chapter 5 support our thinking for utilizing ASes communities under such weighted router level topology to help router-to-AS mapping, the best accuracy rate increase to 82.62%.

Our mainly contributions lie in three parts:

1. We construct a weighted router level topology by using original AS information of IP interfaces and router-pairs similarities simultaneously;
2. We propose a fast hierarchy clustering which time and space complex are both linear, which is capable of finding ASes communities;
3. We demonstrate our methodology using community discovery upon weighted router level topology can lead to a drastically increase of accuracy rate.

Section II we give a brief overview of our methodology, then we discuss some key issues in weighted router topology construction and router communities discovery in section III. Section IV introduces datasets used for constructing global router topology and for validation, then introduces baseline methods for comparing. In Section V, we compare our router-to-AS mapping method with other base methods, while Section VI concludes the paper.

## II. FRAMWORK OF WEIGHTED ROUTER TOPOLOGY CONSTRUCTION AND ROUTER-TO-AS MAPPING

In weighted graph, edges' weights carry more information than topology itself, and community discovery can leverage these weights to get better performance than using topology information alone. In order to construct weighted router topology, we combine IP port information with router ego-network topology together, to give weights between router pairs. As shown in Fig. 1, weighted router topology construction mainly contains three parts. In this section, we give a brief introductions of how weighed router topology can be constructed, and key issues are described in section III.

### *A. Weights Obtaining*

We use IP-AS mapping to find out which AS router-pair may communicate though, because there may have more than one IP port that two routers can communicate with, then AS frequency must be considered. In here, we use "Info Transferring Weight" to describe this frequency: the higher the current AS frequency is, the bigger the weight become. Secondly, we use "Router-Pairs Similarity Weight" to describe router-pairs similarities based on Node Similarity Measurement from Network Science area.

### *B. Weights Fusion*

Because "Info Transferring Weight" and "Router-Pairs Similarity Weight" are irrelevant, in here we use basic operator to combine these information together for simpleness as shown in section III. But thanks to expansibility of our framework, we can also use more complicate methods just as some optimization methods to combine these information together, which is not included in this paper.

### *C. Weighted Community Discovery*

Because AS number is given in the first place, we have to use some community discovery techs such as Hierarchy Clustering which can control output community numbers for avoiding the inequality between AS number and community number. As shown in section V, Hierarchy Clustering method works perfectly in generating community number, and for massive datasets, this kind of method also have good robustness.

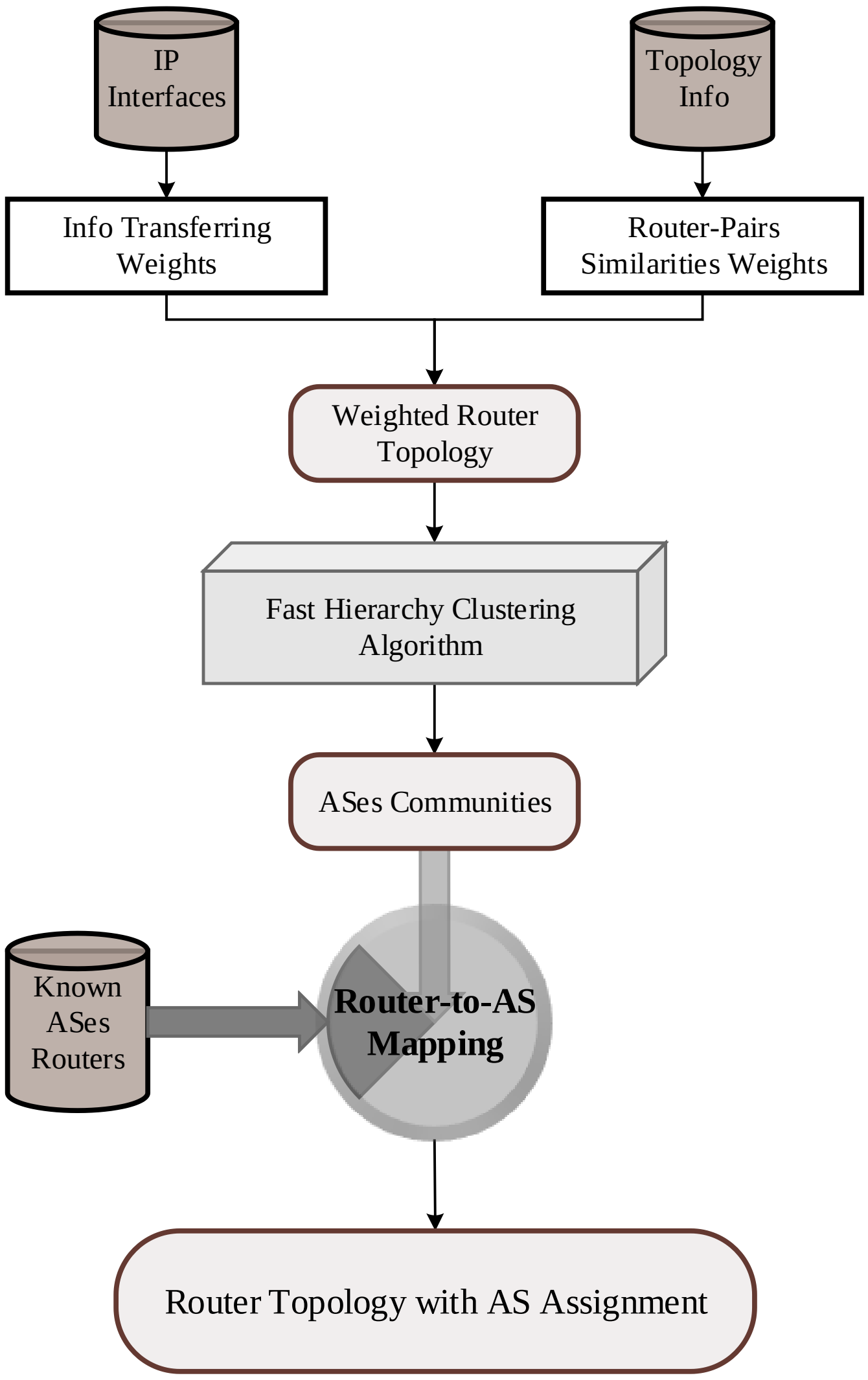

Fig. 1. router-to-AS Mapping Framework from Community Discovery

## III. KEY ISSUES ABOUT OUR METHODOLOGY

As is known to all, there are two distinguishing features for router level topology. First and foremost, a router should be claimed by one and only one AS, which indicate that ASes communities in router network should be non-overlapping; second, since router level is the lower level, which means node size within such topology is generally tremendous. Focus on these two features, we intend to use hierarchy clustering which time and space complex are both linear to do non-overlapping communities discovery on our weighted router level topology. In conclusion, core steps in our methodology are: 1).combining router-pair IP port information with router-pair similarity together to generate weighted topology; 2).using Hierarchy Clustering to do community discovery based on this weighted adjacent matrix on such network; 3).do router-to-AS mapping according to these communities.

In this section, we discuss every key issue in these steps.

### A. Info Transferring Weights Obtaining

As Information transferring procedure between router-pairs through IP ports, using IP-AS mapping we can get which AS or ASes may router-pairs communicate through, in here we use AS frequency to quantify weight between router-pair. Take Fig. 2 as an example, blocks and lines represent Routers and IP ports, AS information above lines represent current IP port's AS number, $l_{AB}$ represents the Info Transferring Weight between RouterA and RouterB. From thisFig. 2, we can tell that RouterA has four IP ports with three of them belong to AS1, and only one belongs to AS2, and information transferring procedure between RouterA and RouterB achieve by through AS1. That means for RouterA, the probability of transferring information through AS1 is 75%, for RouterB, this probability drops to$1/3$ as there are only one port belongs to AS1. So the information transferring weight between RouterA and RouterB is$75\% \times 1 / 3 = 25\%$. For general speaking, we can use Equation (1) to calculate router-pair information transferring weight $W_p$.

$$W_p(A,B) = \begin{cases} \sum_{i,j\in\{AS(A),AS(B)\}} w_i(A) \times w_j(B), if\ adj(A,B) = 1 \\ 0, if\ adj(A,B) = 0\ or\ i \neq j \end{cases} \quad (1)$$

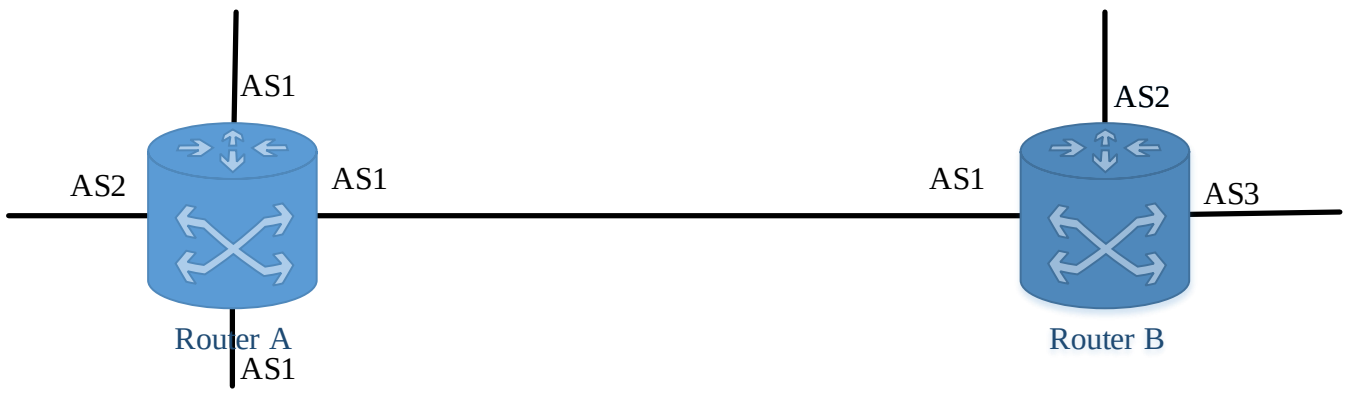

Fig. 2. Toy example of Router-Pairs AS Info

### B. Router-Pairs Similarity Weights Obtaining

In network science area, there are two major metrics to quantify node-pair similarity: 1). Local Similarity Metric: **Only** use node-pair neighbors topologies to depict similarity, such as Jaccard metric; 2). Global Similarity Metric: not only just use neighbors topologies, but also use the whole network topology to depict similarity. From common sense, global similarity metric may perform better than local ones because of containing more information, but for routers network, the communication between a router-pair only involves in these two routers, which has nothing to do with other ones, so in here local similarity metric is good enough for depicting router-pair relationship. But there is a weakness in Local Similarity Metric, as shown in Fig. 3, for example, using Jaccard Metric to quantify node-pair weights, original Jaccard Metric can be defined by Equation (2) where $\Gamma(A)$stands for RouterA's neighbors, then we can tell that $\Gamma(A) \cap \Gamma(B) = \Gamma(B) \cap \Gamma(C) = \phi$, lead to $Jaccard(A,B) = Jaccard(B,C) = 0$, this "ZERO similarity" stands for no edge between RouterA~C, which go against Fig. 3. In order to avoid this failure, we introduce Generalized Neighbors$\Gamma^+(A)$ defined by Equation (3) to update 10 Local Similarity Metric proposed by Zhou T[9], all these updating equations are presenting in TABLE I. In addition, meanings and comparing details are not included in this paper for[9] has a complete introduction and discussion.

$$Jaccard(A,B) = \frac{|\Gamma(A) \cap \Gamma(B)|}{|\Gamma(A) \cup \Gamma(B)|} \quad (2)$$

$$\Gamma^+(A) = \Gamma(A) \cup \{A\} \quad (3)$$

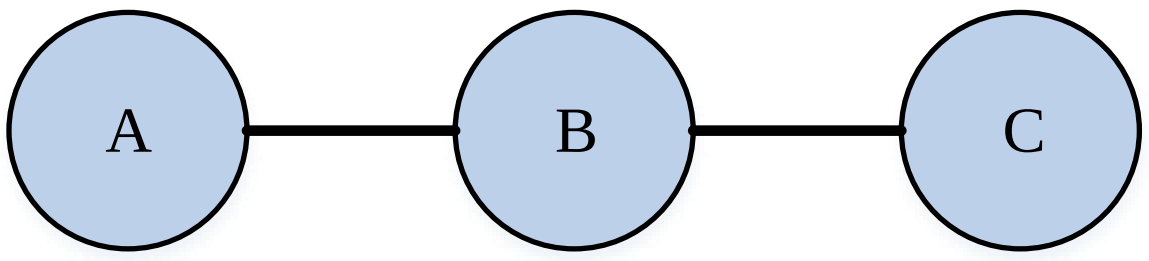

Fig. 3. Toy Example of Jaccard Similarity Failure, with $Jaccard(A,B) = Jaccard(B,C) = 0; Jaccard^+(A,B) = Jaccard^+(B,C) = {}^2/_3$

TABLE I. 10 KINDS OF LOCAL BASED NODE-PAIR SIMILARITY METRICS

| **Metrics** | **Similarity Metric Definition** |
|---|---|
| CN+ | $\|\Gamma^+(A) \cap \Gamma^+(B)\|$ |
| Salton+ | $\|\Gamma^+(A) \cap \Gamma^+(B)\|/\sqrt{K(A) \times K(B)}$ |
| Jaccard+ | $\|\Gamma^+(A) \cap \Gamma^+(B)\|/\|\Gamma^+(A) \cup \Gamma^+(B)\|$ |
| Sorenson+ | $2\|\Gamma^+(A) \cap \Gamma^+(B)\|/K(A) + K(B)$ |
| HPI+ | $\|\Gamma^+(A) \cap \Gamma^+(B)\|/\min(K(A),K(B))$ |
| HDI+ | $\|\Gamma^+(A) \cap \Gamma^+(B)\|/\max(K(A),K(B))$ |
| LHN-I+ | $\|\Gamma^+(A) \cap \Gamma^+(B)\|/K(A) \times K(B)$ |
| PA | $K(A) \times K(B)$ |
| AA+ | $\sum_{\in\Gamma^+(A)\cap\Gamma^+(B)} {}^1/_{\log K(Z)}$ |
| RA+ | $\sum_{Z\in\Gamma^+(A)\cap\Gamma^+(B)} {}^1/_{K(Z)}$ |

### C. Weighted Topology Generation

After obtaining Info Transferring Weights and Router-Pairs Similarity Weights, we have to consider how to generate weighted topology based on these two independent ones. Because there is no theory or previous research results to take examples by, in here we choose four basic operators to do combination, in below, we give our thoughts on why using such operator to generate weighted topology.

- Operator "**Plus**": this operator means Info Transferring Weights and Router-Pairs Similarity Weights take the same importance when we do combination to generate weighted topology.
- Operator "**Times**": because Times means do independent observation on target system, in here we use this operator to depict independence.
- Operator "**Max&Min**": the reason why we introduce these operators is to eliminate average or relative error during "Plus" or "Times" process.

### D. Fast Hierarchy Clustering

Ordinary hierarchy clustering has two core steps to do clustering, first is to do node fusion to form a Dendrogram based on node-pair weights, and second is to use some cutting edge method to cut this Dendrogram to obtain communities. Take Fig. 4as an example, $W_{adj} = (w(i,j))_{N\times N}$ stands for weighted Adjacent Matrix of the Original Network by using Jaccard

Similarity Metric to depict node-pair similarity, then do node-pair fusion to form Dendrogram of such network, then using Modularity Maximum as Cutting Edge Method to find out Communities. From the best cut showing in this Fig. 4, we can easily divide original network into two non-overlapping communities {1,2,3}and{4,5,6}.Take this example in router-to-AS mapping, we can say that if in community one node 1 and node 2 belongs to AS1, then we can say that node 3 may also belongs to AS1. But this simple method may fail when facing massive dataset because for there are $N$ nodes in a Dendrogram, and fusion process can only merge two node together at one time which lead to the Dendrogram height is also$N$. Then we can easily tell that this ordinary hierarchy clustering suffers a high space complex $O(N^2)$. In general speaking, the reason why ordinary hierarchy clustering has such a high space complex is that the algorithm must consider the whole topology at the same time, that means it has to utilize global information to generate the Dendrogram, and this "GLOBAL" leads to consume large amount of memory space. For example, the global router topology described below has more than 2M nodes, then using ordinary clustering method may spend roughly 2M×2M=4K G to store this massive Dendrogram. It is mission impossible for a standard PC which memory space is often 4~8G. So in below, we introduce a new clustering method – "Fast Hierarchy Clustering" to solve this problem.

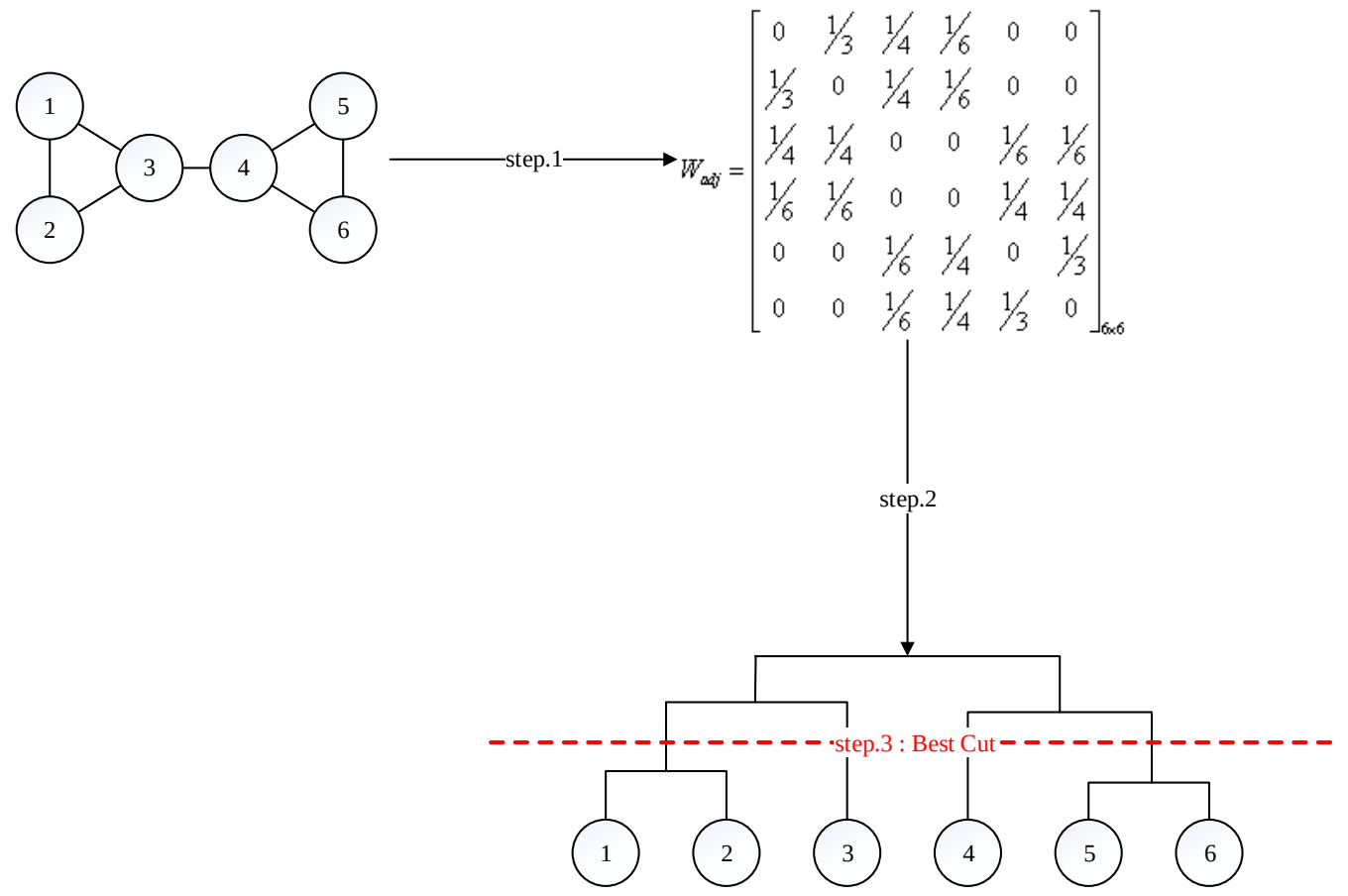

Fig. 4. Toy Example of Ordinary Hierarchy Clustering

As is known to all, the main mechanism of general hierarchy clustering is to aggregate two nodes together that share the same weight, and will never be apart once be aggregated. Leveraging these two characters, a novel algorithm that only relates to the local information of this router's ego network came into our mind. In ordinary hierarchy clustering, we can only do community discovery unless obtaining the whole Dendrogram in the first place, but if we only wish to find which community that a target node belongs to, there is no need for the whole Dendrogram, instead, we can only use this target node's ego network to do so. Take Fig. 5as example, Arabic numerals upon edges means node-pair weight, if we want to find which community that node A belongs to, firstly we can extract node A's ego network, then do aggregation based node A's neighbors' weights. Instead of generating the whole Dendrogram, if we take "**NODE A**" as the aggregating target, then the aggregation of this node become a local problem that is to find another node (node B) in node A's ego network who share the highest weight with node A. And if we do this mechanism iteratively, we can also build a Dendrogram the same as using ordinary method.

After constructing the local Dendrogram, how to identify the community information of every unknown node becomes the second problem. First of all, we use routers which all ports have the same AS number as ground truth; secondly, we combine local Dendrogram and Markov mechanism to assign unknown routers AS number. That means if we have an unknown router A as a target router, then the most probable AS number of router A is the nearest neighbor (maybe router B) of router A in Dendrogram, and if router B's AS number is still unknown, then make router B as the new target, use the same way to find this new target (router B)'s most probable AS number. Using this mechanism recursively, then we can assign first router A's most probable AS number.

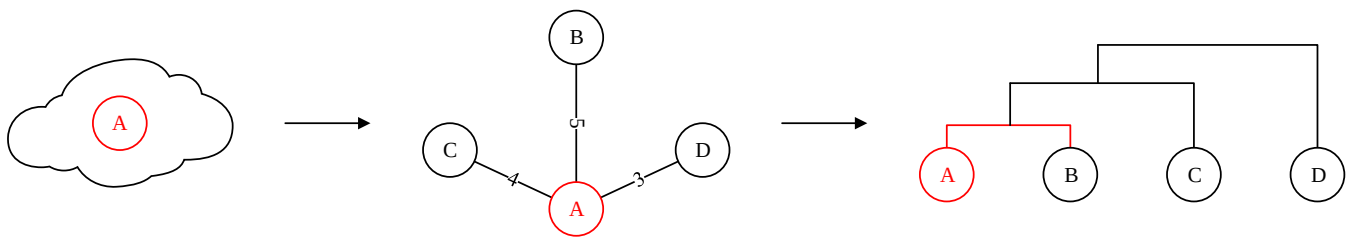

Fig. 5. Toy Example of Local Node Clustering

One concern with our methods is that if Sub Nodes community became bigger and bigger to lead tonone convergence problem. In here we let AS1as the current mapping target AS, $T_1$ as the sub tree which all nodes belongs toAS*, $N(T_1)$ as the number of nodes in $T_1$, and the height of $T_1$is $H_1$. For all nodes in $T_1$, if there are another AS2 beneath $T_1$, which height and number of nodes is $H_2$and $N(T_2)$. Two cases may occur: 1). $H_1 \geq H_2$; 2). $H_1 < H_2$. For case 1, because in the Dendrogram, more height means more nodes will lead to $N(T_1) \geq N(T_2)$naturally, means that when $T_1 = T_2$, the node number in the subtrees will decrease, which can be told that in this case, our method must converge; For case 2, $H_1 < H_2$can infer to $N(T_1) < N(T_2)$, which means all nodes in $T_1$ may belong to AS1, which also lead to a convergence for our method. Hence, our method-Fast Hierarchy Method will not suffer none convergence problem.

## IV. Experiment Study

In this section, we describe the data we collected for experiments and the baseline methods for comparison.

### A. *Data sets Introduction*

We shall examine how our algorithm behaves on global router-level topology. Two datasets are collected: one from CAIDA [17] ITDK project for generating the global router-level topology; and the other from PeeringDB[18] for verifying the Router AS assignment.

*1) CAIDA Router-Level Data for Generating Global Router Topology*

CAIDA (Center for Applied Internet Data Analysis)'s Macroscopic Internet Topology Data Kit (ITDK) is a project that to find all global router-level topology. The ITDK contains data about connectivity and routing gathered from a large cross-section of the global Internet [19]. At present (we are using the router-level topology data that collected in 2012-07, because it is the newest data that we can get access into), this ITDK release consists of four parts: 1) two related router-level topologies, 2) router-to-AS assignments, 3) geographic location of each router, and 4) DNS look ups of all observed IP addresses. In our study, we use part 1 for generating the global router-level topology and part 4 for assigning every router's ports' IP addresses. Because there are 34935241 routers that ITDK contain in 2012-07 with large amount of them have only one port, for simplicity and accuracy, we only maintain routers which have more than one port (port number≥ 2), this deleting one port router processing leads to only 2520154 routers and 19291581edges remaining. Based on these routes and their edges, we construct the global router-level topology and mapping all routers' port IP address to relate AS number using IP alias analysis.

*2) PeeringDB IP-to-AS ground truth Data for Validation*

PeeringDB is a database of networks that are interested in peering [20]. In this database, we can find the true AS that an IP may belong to. After crawling on PeeringDB, we collected 5210 pairs of IP-to-AS correspondences on 2012-02-05. Because our method is mapping router to AS directly, that means we have to first map these IP to Routers manually and then assign related AS to them. We map IP to Router and Router to AS in 3 steps: step.1 is to find routers in ITDK with port IP addresses contains IP in ground truth, after this process, there are only 617 routers that satisfy the condition; in step.2 we look up router-level topology generating before to examine if these routers were isolated nodes or not, and we find there are 3 of them is solo nodes, so we take them out; step.3 is to take out current router's port IP which have no known AS information, for example, if router *A* has 3 IP where AS refers to AS1, AS2 and AS3 according to IP alias analysis, but router *A*'s AS of ground truth is AS4 according to step.1, that means we can never use AS1~AS3 to infer AS4, so we have to take router *A* out with lack of other information, during this process, we find there are 125 of 617 routers suffer this situation, and we have to take them out too. After all of these three steps, there are 489 routers remaining for validation. So we take these 489 routers as ground truth.

## *B. Baseline Methods*

We apply our method to do router-to-AS mapping based on router-level topology data, and use ground truth to validate them, and we also compare our method to some representative router AS Clustering method.

*1) CAIDA Method: Election Tech +Degree Tech*

Huffaker propose a method that can find router's current AS only based on its ports [5]. This method is mainly based on Election tech, and Degree tech can be only used when Election has failed to assign current AS to the router. First of all, we take Fig. 6 to see how Election Tech works: for Router A has three ports which AS is assigned as AS1, AS1 and AS2, then Router A's AS may assign as AS1 because two of all are AS1;Secondly,we take Fig. 7 to see how Degree Tech works: for Router B had three ports AS number as AS1, AS2 and AS2, in order to get what AS number Router B belongs to, the degree tech first generate an AS-level topology by assuming full-mesh connectivity among ASes from each router's AS frequency matrix(step.1), then use this topology graph to get each AS's degree(step.2). According to the basic knowledge "the AS most likely to be the customer AS, based on similar intuition as the Customer heuristic [5]", degree tech assigns Router B to AS1 because of the smallest degree in AS topology.

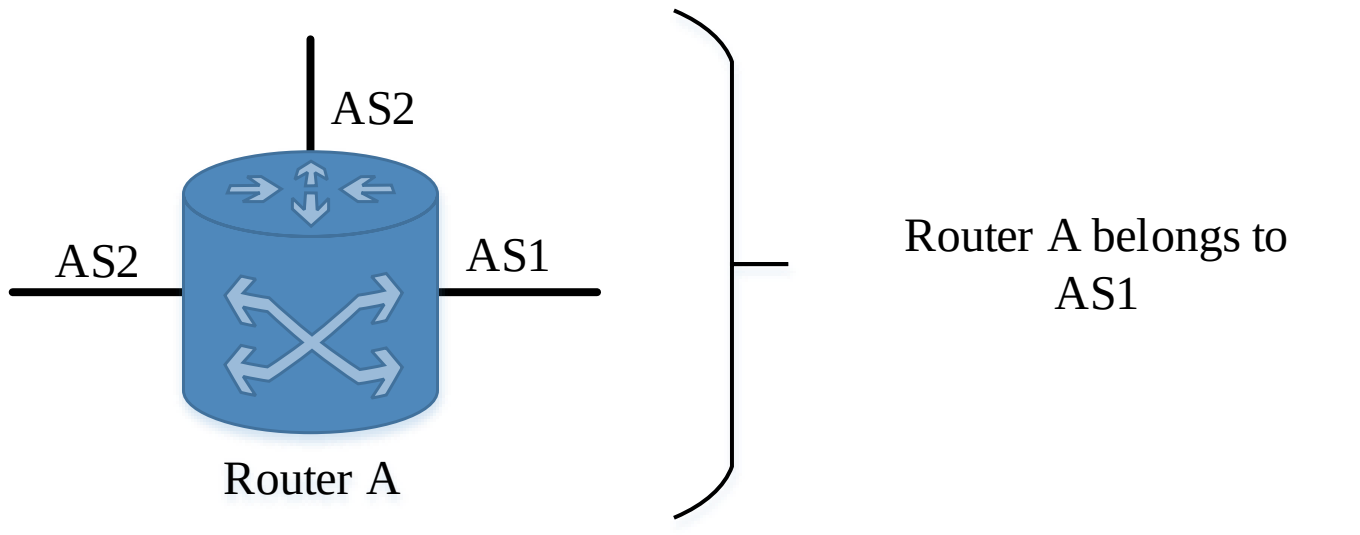

Fig. 6. Toy Example of Election Tech

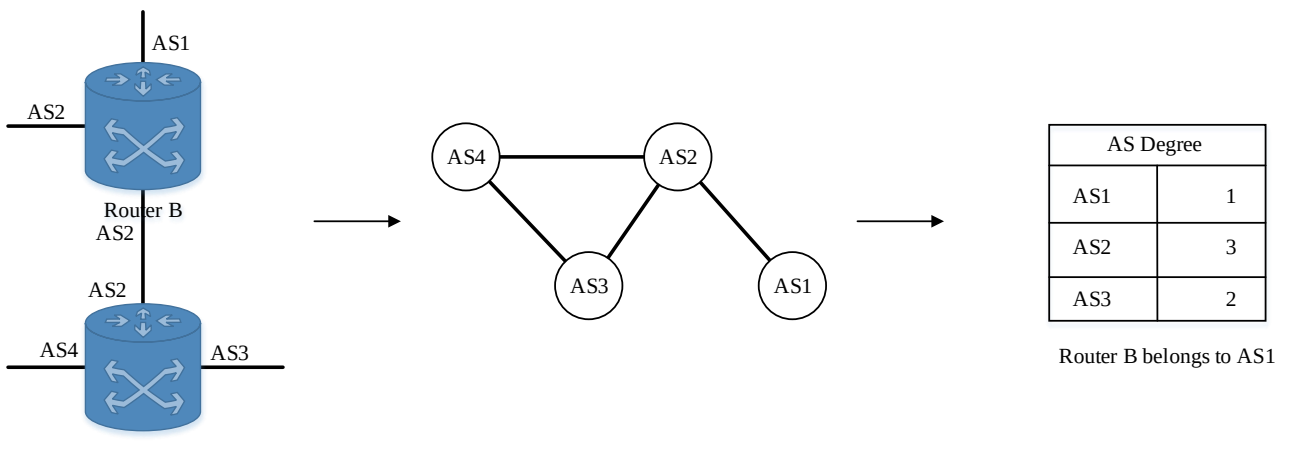

Fig. 7. Toy Example of Degree Tech

*2) Hierarchy Clustering based on TOPOLOGYONLY& Hierarchy Clustering base on PORTONLY*

In our baseline methods, not only do we use the classical and simplest router AS assignment method from CAIDA as our base line method but also for integrality, we test our hierarchy clustering based on topology and port number only, that means we only construct our weighted router adjacent matrix with only topology information or only port number information. In below, we can see that in some metric which we have discussed before, even using partial information from topology, the clustering results still better performed than classical method from CAIDA.

# V. Experiment Results

In this section, we mainly focus on how our router-to-AS Mapping method and other baseline methods behave on global router-level topology, as discussed above, we use PeeringDB data as ground truth, and apply clustering method on global topology based on CAIDA ITDK project. It is easily to tell that the more routers from ITDK can be assigned as real AS, the more accurate the algorithm has. After comparing performance of different router-to-AS Mapping methods, we try to explain the reason why our router-to-AS Mapping method under such framework can be out performed.

First of all, we apply CAIDA method (Election + Degree) to the ground truth, and find there are only 320 routers are correct. That means the CAIDA method has only 65.44% accuracy rate. Secondly, we use Fast Hierarchy Clustering method to do router AS assignment. As described before, there are three kinds of information that we can leverage to construct Weighted Router Topology network in Fast Hierarchy Clustering, so we do these ways separately. Way.1: We only use node port information to generate weight between two nodes; Way.2: We only use original node similarity to generate weight between two nodes; Way.3: We only use modified node similarity to generate weight between two nodes; Way.4: We combine original node similarity and router's port information to generate weight between two nodes; Way.5: We combine modified node similarity and router's port information to generate weight between two nodes.

For Way.1, because of considering router's port information alone, we find that only 336 routers can be assigned as real AS, so the accuracy rate is 68.71%, which is a little more than CAIDA method. For Way.2~5, because there are 10 metrics to calculate node pair similarity as described before, so we use Fig. 8 to demonstrate the accuracy rate for each metric in each way. In this Fig. 8, horizontal axis stands for 10 metrics, vertical axis stands for accuracy rate ranging from 0% to 90% because the highest one is 82.62%; as is shown on legend, each color stands for each way to calculate accuracy rates.

From Way.2, we can see that if only using Original Node Similarity Metric, then accuracy rates are all lower than CAIDA method 65.44% except RA Metric, and even for RA Metric, the accuracy rate only increase to 67.89%, which is also lower than ONLY use Port Information 68.71%. This experiment gives another proof of the incompleteness of using topology information alone without port information from original router network.

Way.3 demonstrates that ONLY using "Modified Node Similarity" gains a tiny increment of accuracy rates, but it also suffers the same incompleteness problem which has already been described in front. It's worth noting that the Decrement of Accuracy Rate while using Jaccard$^+$ Metric, we think the main reason of this Decrement is because, there are lots of router connections like Fig. 3in Router Global Network Topology, as there are only one connection between router A, B and C. If using Original Node Similarity Metric, there will be ZERO similarity between these three routers, which lead to non-combination in Fast Hierarchy Clustering, but Modified Node Similarity Metric will never assign the ZERO similarity between two routers, this character brings in some unexpected noise during Fast Hierarchy Clustering Process, and that may lead to the decrement of accuracy rate using Jaccard$^+$ Metric.

In Way.4, we combine Original Node Similarity and Port Information together, from the Fig. 8 shows that the accuracy rate is better than only using original node similarity asWay.2, this proves another time that only use one information may not be good enough for router-to-AS Mapping.

Combine Modified Node Similarity with Port information gets the **best accuracy rates** showing in **Way.5**. We can see that except CN$^+$, PA$^+$, AA+ metrics, all other metrics perform better than 65.44%, and using ***RA$^+$ metric, accuracy rate can reach 82.62%.***

After do analysis for each way, here we also try to explain why different similarity calculations get different results. These 10 similarity metrics can be divided into three ways according to different emphasis of features: 1). Common Neighbor as main Feature (Metric 1~4,7); 2). Hub Node as main Feature (Metric 5,6); 3). Resource Allocation Theory (Metric 9,10). Here we discuss these three kinds of features separately.

First of all, from CN$^+$ metric and metric2~7 according to different normalization methods, LHN-I$^+$ gets the highest accuracy rate 70.14%, and CN$^+$ gets the lowest 39.26%. We think the reason why CN$^+$ gets the lowest accuracy rate lies in lacking of normalization process, that lead to CN$^+$ metric's range beyond [0,1], but the Port Information range limits in [0,1], so the low accuracy should be normal.

Secondly, the high accuracy rate of LHN-I$^+$ metric represents that for global router network topology, the similarity metric of router-pairs have a reverse ratio with router's degree, but a direct ratio with router's neighbors numbers; Secondly, for HPI$^+$ and HDI$^+$ metric, it can be told that accuracy rate of HPI$^+$ perform better than HDI$^+$, that indicates for routers, they are also more willing to connect to other routers who have higher degrees, this phenomenon can be also found in Complex network analysis.

Lastly, according Resource Allocation Theory, the only difference of RA$^+$ metric and AA$^+$ metric is the weight increment pattern $1 / K(Z)$ or $1 / (\log K(Z))$. From Fig. 8, we can see that RA metric performs better than AA, which is another powerful proof of Occam's razor principle: Entities should not be multiplied unnecessarily.

The above analysis of the experiments shows that ***it is best to use Modified Node Similarity and Port Information to depict router-pairs similarity features***, this means when doing community discovery in some particular fields, we must combine original community discovery with some specific features from this field to create an adaptable algorithm in such field. In this paper, we adopt this guiding ideology to come up with a new framework to deal with router-to-AS Mapping problem, which combine Port Information feature from original router-to-AS Mapping tech with Modified Node Similarity Metric from community discovery. Through the discussion of the above, these two features are indispensable.

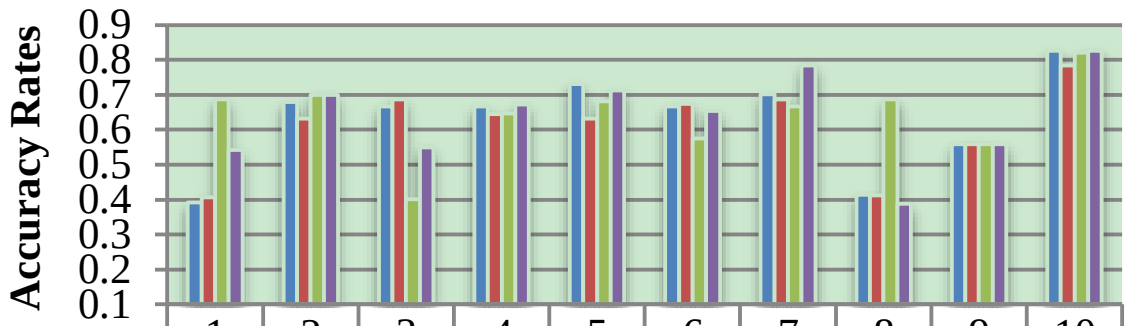

| | 1 | 2 | 3 | 4 | 5 | 6 | 7 | 8 | 9 | 10 |
|---|---|---|---|---|---|---|---|---|---|---|
| PLUS | .393 | .679 | .667 | .667 | .730 | .667 | .701 | .414 | .558 | .826 |
| MAX | .407 | .632 | .687 | .644 | .632 | .675 | .687 | .413 | .558 | .783 |
| MIN | .687 | .699 | .401 | .646 | .681 | .575 | .667 | .687 | .558 | .820 |
| TIMES | .542 | .699 | .550 | .673 | .712 | .654 | .783 | .389 | .558 | .826 |

**Node Similarity Metric**

Fig. 8. Accuracy Rates Comparing

In addition, it is clearly to find that when we do fast hierarchy clustering on routers, we not only use routers that have only one port AS number as ground truth, but also use routers that have been assigned before, that may lead to a problem: "Do these routers' sequence have influence on the result?" In order to answer this question, we randomly resort these routers' sequences for 10 times, and do our Clustering method on each of them. TABLE II. demonstrates the average accuracy rate and standard error, from this table we can tell that our Fast Hierarchy Clustering method has a strong robust and stability for almost every standard error is less than 0.01.

TABLE II. ACCURACY RATE AND STANDARD ERROR BASED ON RANDOM SORTED ROUTERS' SEQUENCES

| Metrics | Average Accuracy Rate | Standard Error |
|---|---|---|
| $CN^+$ | 38.79% | 0.0053 |
| $Salton^+$ | 67.95% | 0.0010 |
| $Jaccard^+$ | 66.69% | 0.0006 |
| $Sorenson^+$ | 66.83% | 0.0012 |
| $HPI^+$ | 73.05% | 0.0008 |
| $HDI^+$ | 66.73% | 0.0009 |
| $LHN\text{-}I^+$ | 69.33% | 0.0000 |
| PA | 39.53% | 0.0103 |
| $AA^+$ | 55.83% | 0.0000 |
| $RA^+$ | **82.62%** | 0.0000 |

At last, there remains one question about how to combine Node Similarity and Port information together. In front, we use "PLUS" to do this combination, but if it weren't the best to combine these two irrelevant information together? In below, we consider "MAX", "MIN" and "Times" these other three operators to combine these two information together. Fig. 9givesthe accuracy rate for each operator. From it, we can see that "**Times+RA**$^+$" and "**Plus+RA**$^+$" share the same highest accuracy rate as**82.62%**, which again proves Port Information and Modified Node Similarity Metric these two features are indispensable.

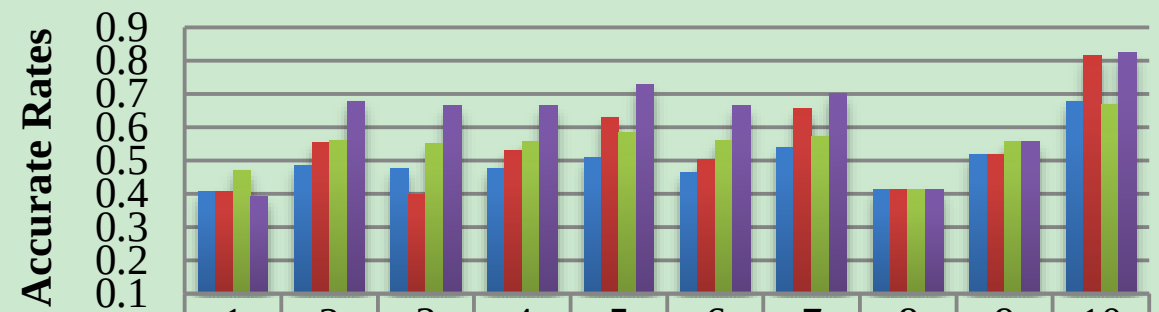

| | 1 | 2 | 3 | 4 | 5 | 6 | 7 | 8 | 9 | 10 |
|---|---|---|---|---|---|---|---|---|---|---|
| Way.2 | .407 | .487 | .477 | .477 | .509 | .464 | .542 | .413 | .519 | .679 |
| Way.3 | .407 | .554 | .399 | .532 | .632 | .503 | .656 | .413 | .519 | .816 |
| Way.4 | .472 | .562 | .550 | .558 | .585 | .560 | .573 | .414 | .558 | .669 |
| Way.5 | .393 | .679 | .667 | .667 | .730 | .667 | .701 | .414 | .558 | .826 |

**Node Similarity Metrics**

Fig. 9. Four Different Operators Comparing

## VI. CONCLUSION

In this paper, we propose a new methodology to do router-to-AS mapping. This methodology mainly combines ordinary router-to-AS mapping techs from pure Internet area with Community Discovery from Network Science to obtain more accurate rate. The most advantage in this work is to show us a way to combine irrelevant information from different research domains together to achieve better performance. In our work, we first update node similarity definition $\Gamma(\cdot)$ to $\Gamma^+(\cdot)$ to avoid unnecessary noise, and then develop a new fast hierarchy clustering method to do router-to-AS mapping under massive dataset (Global Router Topology). The results shows that, using $RA^+$ metric and AS frequency to quantify node-pair similarity weights and IP port information weights to generate weighted topology, with using our Fast Hierarchy Clustering to assign AS to routers, the accuracy rate can reach to 82.62%, that is far more better than "Election + Degree" method which can only reach 65.44% introduced by CAIDA. At the same time, we compare four basic operators "Plus", "Times", "Max", "Min" to do weight fusion to uncover how these two irrelevant weights effect on each other, the result shows that both "Plus" and "Times" can get the best accuracy rate 82.62%, which means these two irrelevant information weighted equally. At last, we also use experimental result to prove that our method under this novel framework of router-to-AS Mapping inspired by Community Discovery has a strong robustness.

## ACKNOWLEDGMENT

Thanks to Yuan Mingkai for his work on processing datasets and Jiang Qing for her knowledge on IP-AS mapping, and this work was supported by National Natural Science Foundation of China (No.61301274).

## REFERENCES

[1] He Y, Siganos G, Faloutsos M, et al. Lord of the Links: A Framework for Discovering Missing Links in the Internet Topology[J]. IEEE/ACM Transactions on Networking, 2009, 17(2):391-404.

[2] Oliveira R, Pei D, Willinger W, et al. The (In)Completeness of the Observed Internet AS-level Structure[J]. Networking IEEE/ACM Transactions on, 2010, 18(1):109 - 122.

[3] Keys K, Hyun Y, Luckie M, et al. Internet-scale IPv4 alias resolution with MIDAR[J]. IEEE/ACM Transactions on Networking, 2013, 21(2):383-399.

[4] Gunes M H, Sarac K. Resolving Anonymous Routers in Internet Topology Measurement Studies[C]// INFOCOM 2008. The 27th Conference on Computer Communications. IEEE. IEEE, 2008:1076-1084.

[5] Rekhter Y, Katz D, Mathis M, et al. Application of the border gateway protocol in the internet[J]. 1990.

[6] Mao Z M, Rexford J, Wang J, et al. Towards an Accurate AS-Level Traceroute Tool[J]. Acm Sigcomm Computer Communication Review, 2003, 33(4):365-378.

[7] Tangmunarunkit H, Doyle J, Govindan R, et al. Does AS Size Determine Degree in AS Topology?[J]. Acm Sigcomm Computer Communication Review, 2001, 31(5):7-10.

[8] Tangmunarunkit H, Govindan R, Shenker S, and Estrin D. The Impact of Routing Policy on Internet Paths. In Proc. IEEE INFOCOM, 2001

[9] Pansiot J J, Mérindol P, Donnet B, et al. Extracting Intra-domain Topology from mrinfo Probing.[C]// Proceedings of the 11th international conference on Passive and active measurement. Springer-Verlag, 2010:81-90.

[10] Huffaker B, Dhamdhere A, Fomenkov M, et al. Toward Topology Dualism: Improving the Accuracy of AS Annotations for Routers[M]// Passive and Active Measurement (PAM). Springer Berlin Heidelberg, 2010:101-110.

[11] Albert R, Barabási A L. Statistical mechanics of complex networks[J]. Reviews of modern physics, 2002, 74(1): 47.

[12] Loe C W, Jensen H J. Comparison of communities detection algorithms for multiplex[J]. Physica A: Statistical Mechanics and its Applications, 2015, 431: 29-45.

[13] Newman M E J. The structure and function of complex networks[J]. SIAM review, 2003, 45(2): 167-256.

[14] Palla G, Derenyi I, Farkas I, et al. Uncovering the overlapping community structure of complex networks in nature and society [J]. Nature, 2005, 435(7043):814-818.

[15] Rekhter Y, Katz D, Mathis M, et al. Application of the border gateway protocol in the internet[J]. 1990.

[16] Zhou T, Lü L, Zhang Y C. Predicting missing links via local information[J]. The European Physical Journal B-Condensed Matter and Complex Systems, 2009, 71(4): 623-630.

[17] http://www.caida.org/home/

[18] https://www.peeringdb.com/private/

[19] ITDK Data:http://data.caida.org/datasets/topology/ark/ipv4/itdk/2012-07/

[20] https://www.peeringdb.com/help/public_faq.php